\documentclass[prl,aps,preprint,floatfix]{revtex4}
\usepackage{amssymb}
\usepackage{epsfig}
\usepackage{graphicx}
\usepackage[centertags]{amsmath}
\usepackage{latexsym}
\usepackage{natbib}
\usepackage{textcomp}
\usepackage{color}

\begin{document}

\begin{center}

\textbf{Novel highly conductive and transparent graphene based conductors}\\

\end{center}

\begin{center}
\textit{Ivan Khrapach, Freddie Withers, Thomas H. Bointon, Dmitry K. Polyushkin, William L. Barnes, Saverio Russo, and Monica F. Craciun*}\\
\end{center}

\begin{center}
Centre for Graphene Science, College of Engineering, Mathematics and Physical Sciences, University of Exeter, Exeter, EX4 4QL, UK\\
\end{center}

\begin{center}
E-mail: m.f.craciun@exeter.ac.uk\\
\end{center}





Future wearable electronics, displays and photovoltaic devices require materials which are mechanically flexible, lightweight and low-cost, in addition to being electrically conductive and optically transparent \cite{REV1,REV2,REV3}. Nowadays indium tin oxide (ITO) is the most wide spread transparent conductor in optoelectronic applications, however the mechanical rigidity of this material limits its use for future flexible devices. In the race to find novel transparent conductors, graphene monolayers and multilayers are the leading candidates as they have the potential to satisfy all future requirements. Graphene, one-atom-thick layer of carbon atoms, is transparent\cite{15transparency}, conducting \cite{24Novoselov,25Novoselov}, bendable \cite{17Roll} and yet one of the strongest known materials \cite{Lee2008}. However, the use of graphene as a truly transparent conductor remains a great challenge because the lowest values of its sheet resistance (R$_{s}$) demonstrated so far are above the values of commercially available ITO (i.e. 10 $\Omega/\Box $ at an optical transmittance Tr=85$\%$ \cite{27limits}). Currently many efforts are concentrated on decreasing the R$_{s}$ of graphene-based materials while maintaining a high Tr, which will allow their potential to be harnessed in optoelectronic applications. To date, the best values of sheet resistance and transmittance found in graphene-based materials are still far from the performances of ITO, with typical values of R$_{s}$=30 $\Omega/\Box $ at Tr=90$\%$ for graphene multilayers \cite{17Roll,multilayers} and R$_{s}$=125 $\Omega/\Box$ at Tr=97.7$\%$ for chemically doped graphene \cite{17Roll,18doping1,19doping2}.

Here we report novel graphene-based transparent conductors with a sheet resistance of 8.8 $\Omega/\Box$ at Tr=84$\%$, a carrier density as high as 8.9$\times10^{14}cm^{-2}$ and a room temperature carrier mean free path as large as $\sim$ 0.6$\mu m$. These materials are obtained by intercalating few-layer graphene (FLG) with ferric chloride (FeCl$_3$) \cite{21Ferrari,23Zhan}. Through a combined study of electrical transport and optical transmission measurements we demonstrate that FeCl$_3$ enhances the electrical conductivity of FLG while leaving these graphene-based materials highly transparent. We also show that FeCl$_3$-FLG are stable in air up to one year, which demonstrates the potential of these materials for industrial production of transparent conductors. The unique combination of record low sheet resistance, high optical transparency and macroscopic room temperature mean free path has not been demonstrated so far in any other doped graphene system, and opens new avenues for graphene-based optoelectronics.

Pristine FLG ranging from two- to five-layers (2L to 5L) were obtained by micromechanical cleavage of natural graphite \cite{24Novoselov} on glass or SiO$_{2}$/Si. The number of layers composing each FLG was determined by optical contrast and Raman spectroscopy (see Supporting Information). The intercalation process with FeCl$_3$ was performed in vacuum with the two-zone vapor transport method \cite{20Dresselhaus}. FeCl$_3$ is sublimated in the lowest temperature zone and it diffuses to the higher temperature zone where the intercalation of FLG takes place (see Experimental section). The structural, electrical and optical characterization is carried out by means of three complementary experimental techniques: low-temperature charge transport, optical transmission and Raman spectroscopy.

Figure 1a shows the Raman spectra of pristine FLG on SiO$_{2}$/Si, with the G-band at 1580 $cm^{-1}$ and the 2D-band at 2700 $cm^{-1}$ \cite{Ferrari2006,Malard2009}. As expected for pristine FLGs, increasing the number of layers results in an increase of the G-band intensity \cite{Koh2011}, whereas the 2D-band acquires a multi-peak structure \cite{Ferrari2006,Malard2009}. The charge transfer from FeCl$_3$ to graphene modifies the Raman spectra of FLGs in two distinctive ways \cite{21Ferrari,22Kim,23Zhan}: an upshift of the G-band and a change of the 2D-band from multi- to single-peak structure, respectively (see Figure 1a). The shift of the G-band to G$_1$=1612 $cm^{-1}$ is a signature of a graphene sheet with only one adjacent FeCl$_3$ layer, whereas the shift to G$_{2}$=1625 $cm^{-1}$ characterizes a graphene sheet sandwiched between two FeCl$_3$ layers \cite{20Dresselhaus,21Ferrari,23Zhan} (see Figure 1a). The frequencies, linewidths and lineshapes of the G$_1$ and G$_{2}$ peaks do not depend on the number of graphene layers which indicates the decoupling of the FLGs into separate monolayers due to the intercalation of FeCl$_3$ between the graphene sheets. This is consistent with the changes in the 2D-band shape and with the Raman studies of other intercalants such as Potassium \cite{Howard,Jung} and Rubidium \cite{Jung}. These observations allow us to identify the structure of intercalated 2L samples as one FeCl$_3$ layer sandwiched between the two graphene sheets. However, the structural determination of thicker FeCl$_3$-FLG cannot rely uniquely on the Raman spectra, it requires complementary knowledge from electrical transport experiments. Direct structural determination for example by X-ray diffraction would be valuable to confirm the findings of Raman and electrical transport measurements, however the small thickness of FeCl$_3$-FLG and the substrate effects make it difficult to apply such technique to these systems.

To characterize the structure of thicker FeCl$_3$-FLG, we study the oscillatory behaviour of the longitudinal magneto-conductance ($G_{xx}$) in a perpendicular magnetic field (i.e. Shubnikov-de Haas oscillations, SdHO) in combination with the Hall resistance ($R_{xy}$). Here we discuss the representative data for an intercalated 5L sample patterned into a Hall bar geometry (see Figure 1c and Experimental section). Figure 1b shows SdHO of G$_{xx}$ as a function of perpendicular magnetic field (B) for different temperatures. It is apparent that for T$<$10K G$_{xx}$ oscillates with two distinct frequencies. For T$>$10K only the lower frequency oscillations are visible. These observations indicate that electrical conduction takes place through parallel gases of charge carriers with distinct densities. Indeed, the Fourier transform of G$_{xx}$(1/B) yields peaks at frequencies $f_{SdHO_1}=1100T$ and $f_{SdHO_2}=55T$ (see Figure 1d), corresponding to charge carrier densities $n_1=(1.07\times10^{14}\pm 5\times10^{11})cm^{-2}$ and $n_2=(5.3\times10^{12}\pm 4\times10^{11})cm^{-2}$ (with $n_i=4ef_{SdHO_i}/h$ \cite{25Novoselov,Kim2005}).

\begin{figure}[!h] 
\centering
\includegraphics[width=1\textwidth]{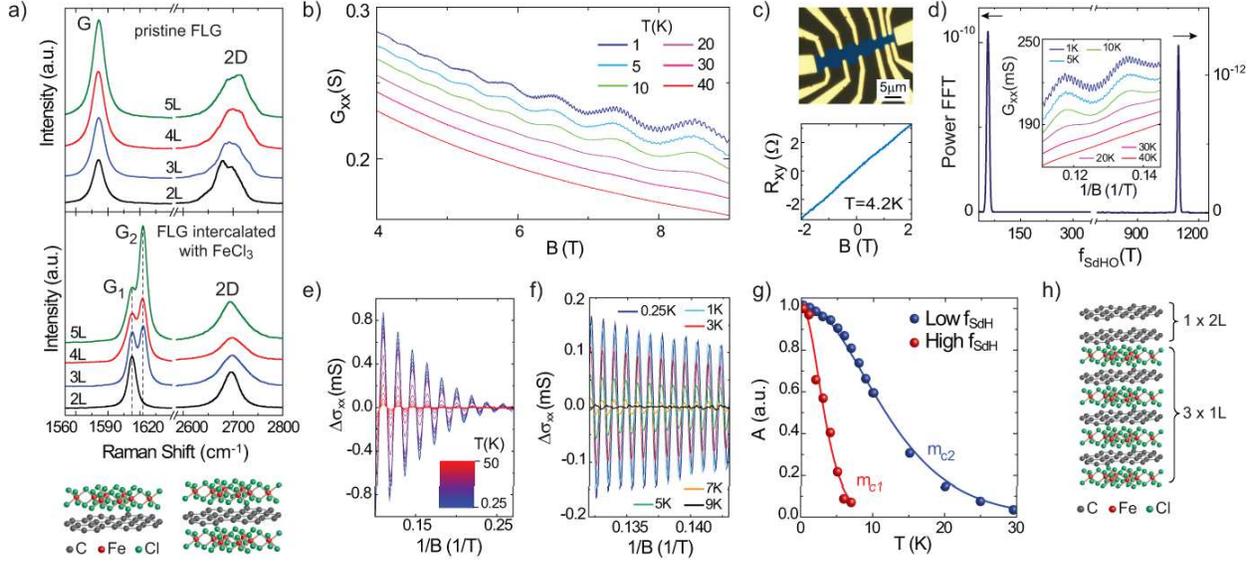}
\noindent{\caption{a) The G and 2D Raman bands of pristine FLG (top) and of FeCl$_3$-FLG (bottom) with different thicknesses ranging from 2L to 5L. The Raman shift of G to G$_1$ and G$_{2}$ stem for a graphene sheet with one or two adjacent FeCl$_3$ layers as shown by the schematic crystal structure. b) Longitudinal conductance ($G_{xx}$) as a function of magnetic field at different temperatures (curves shifted for clarity). c) Top panel: optical microscope image of a Hall bar device. Bottom panel: Hall resistance ($R_{xy}$) as function of magnetic field. d) Fourier transform of G$_{xx}$(1/B) with peaks at frequencies $f_{SdH}^{(1)}=1100T$ and $f_{SdH}^{(2)}=55T$. The inset shows G$_{xx}$ as a function of inverse magnetic field at different temperatures (curves shifted for clarity). Panels e) and f), respectively, show the low- and high- frequency magneto-conductivity oscillations vs 1/B extracted from the measurements in b) (see Experimental section). g) Temperature decay of the amplitude (A) of $\Delta \sigma_{xx}$ oscillations at B=$6.2T$. The amplitudes are normalized to their values at T=$0.25K$. The continuous lines are fits to $A(T)/A(0.25)$ with the cyclotron mass $m_c$ as the only fitting parameter. h) Schematic crystal structure of a 5L FeCl$_3$-FLG in which electrical transport takes place through four parallel conductive planes, one with bilayer character and three with monolayer character.} \label{fig1}}
\end{figure}

The temperature dependence of the magneto-conductivity oscillations allows us to determine the cyclotron mass of the charge carriers in these parallel gases. Figures 1e and f show the low- and high-frequency magneto-conductivity oscillations for different temperatures (see Experimental section). In all cases, the temperature decay of the amplitude is well described by $A(T)\propto T/sinh(2\pi^2k_BTm_c/\hbar eB)$ (see Figure 1g), with cyclotron masses $m_{c_1}=(0.25$\textpm$0.05)m_e$ and $m_{c_2}=(0.08\pm0.001)m_e$ for the high- and low-frequency oscillations, respectively. These values correspond to the expected values of cyclotron mass for massless Dirac fermions $m_c=\sqrt{h^2n/4\pi v_F^2}=0.21m_e$ at $n_1$ and for chiral massive charge carriers of bilayer graphene $m_c=\sqrt{\hbar^2v_F^2\pi n+(\gamma/2)^2}/v_F^2=0.084m_e$ at $n_2$ (with $v_F=10^{6}$m/s the Fermi velocity \cite{25Novoselov} and $\gamma$ the interlayer hopping energy \cite{20Dresselhaus}), see Supporting Information. Therefore, intercalation of FeCl$_{3}$ decouples the stacked 5L graphene into parallel gases of massless (1L) and massive (2L) charge carriers.

The charge carrier type (electrons or holes) and the number of parallel gases present in FeCl$_3$-FLG is readily identified by correlating SdHO to the Hall resistance measurements. The linear dependence of R$_{xy}$(B) with positive slope identifies charge carriers as holes, with a Hall carrier density $n_{H}=B/(eR_{xy})$=$3\times10^{14}cm^{-2}$ (see Figure 1c). Since the total carrier density $n_{tot}=\sum_i n_i$ should be higher than $n_{H}=(\sum_i n_i\mu_i)^2/\sum_i n_i\mu_i^2$ (with $n_{i}$ and $\mu_i$ the carrier density and mobility of each hole gas) \cite{26Davies}, only a minimum of three parallel hole gases with $n_1=1.07\times10^{14}cm^{-2}$ can explain the estimated value of $n_{H}$, i.e. $3\times n_1+n_2\geq n_{H}$ (see Supporting Information). Therefore, the electrical transport characterization demonstrates the presence of four parallel hole gases, of which one with bilayer character (and density $n_2$) and three with monolayer character (each with density $n_1$). These findings are confirmed by the Raman spectra taken after the device fabrication showing the presence of pristine G, G$_1$ and G$_{2}$ peaks (see Supporting Information). A schematic of this crystal structure is illustrated in Figure 1h. The bilayer gas is likely to be caused by the first two layers of the stacking which have been de-intercalated due to rinsing in acetone during lift-off \cite{22Kim}. The bottom part of the stacking has a  per-layer doping of $n_1=1.07\times10^{14}cm^{-2}$ and the stoichiometry of stage-1 FeCl$_3$ graphite intercalation compounds (i.e. where each graphene layer is sandwiched by two FeCl$_3$ layers) \cite{20Dresselhaus}. In total we have investigated electrical transport and Raman spectroscopy in more than 10 intercalated 5L samples and in all cases we confirmed the structure reported in Figure 1h.

A remarkable property of FeCl$_3$-FLGs shown by the electrical transport characterization is that intercalated materials thicker than 3L invariably exhibit a very low-sheet resistance, which is essential for their use as electrical conductors. We find a room temperature value of R$_{s} = 8.8 \Omega/\Box$ in 5L intercalated FLGs. The 4L intercalated FLGs typically exhibit higher sheet resistance values than the 5L intercalated samples with a similar crystal structure (see Supporting Information for a comparison between the electrical properties and Raman spectra of several 4L and 5L intercalated FLGs). Furthermore, the sheet resistance of FeCl$_3$-FLGs thicker than 2L decreases when lowering the temperature as expected for metallic conduction (see Figure 2a). Contrary to intercalated samples, pristine FLGs always have a higher R$_{s}$ ($>$120 $\Omega/\Box$) and they exhibit non-metallic behaviour as a function of temperature \cite{24Novoselov,Craciun2009,Avouris2009} (see Figure 2b). This suggests that the origin of the low values of R$_{s}$ and the metallic nature of the conduction are consequences of intercalation with FeCl$_3$. FeCl$_3$-FLGs thinner than 3L suffer of partial de-intercalation during the device fabrication, which results in a non-metallic behaviour similar to pristine FLGs and in higher R$_{s}$ values (see Figure 2a).

\begin{figure}[!h] 
\centering
\includegraphics[width=0.8\textwidth]{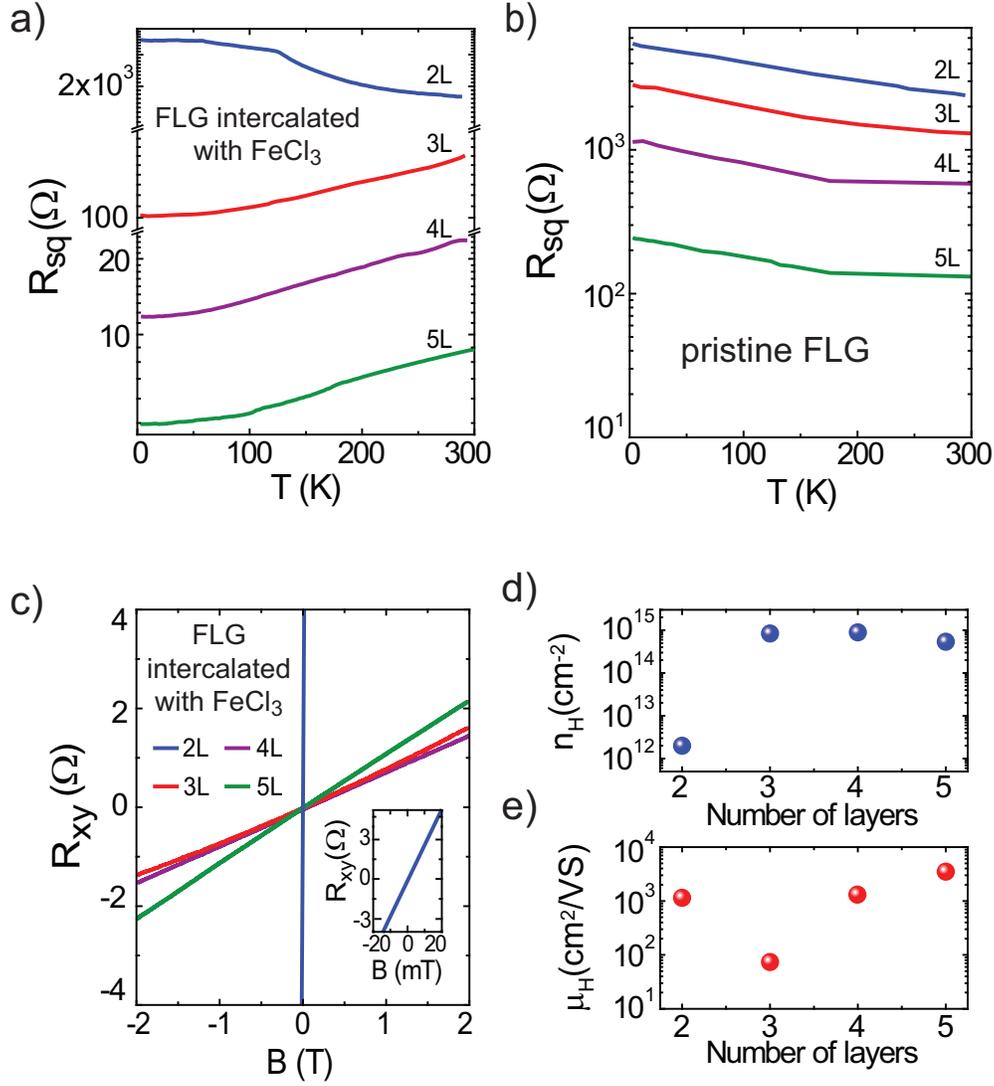}
\noindent{\caption{a) Temperature dependence of the square resistance for FeCl$_3$-FLG of different thicknesses. b) Square resistance for pristine FLG of different thicknesses as function of temperature. These devices are fabricated on SiO$_2$/Si substrates and the highly-doped Si substrate is used as a gate to adjust the Fermi level to the charge neutrality of the system. c) Hall resistance of FeCl$_3$-FLG as a function of magnetic field. The inset shows the data for the bilayer sample on a smaller B scale. Panels d) and e) show the carrier density and mobility for FeCl$_3$-FLG as a function of the number of graphene layers.} \label{fig2}}
\end{figure}

The low values of R$_{s}$ characterizing thick FeCl$_3$-FLG are accompanied by an extremely high charge density. Indeed, the Hall coefficient measurements reveal that $n_{H}$ ranges from 3$\times10^{14}cm^{-2}$ to 8.9$\times10^{14}cm^{-2}$, depending on the number of layers (see Figures 2c and d). These charge densities exceeds even the highest values demonstrated so far by liquid electrolyte \cite{Efetov} or ionic \cite{Ye} gating. The corresponding charge carrier mobility for the 4L and 5L samples typically
ranges from $\mu_{H}=1540 cm^2/Vs$ to $\mu_{H}=3650 cm^2/Vs$ ($\mu_{H}=1/(n_{H}e\rho_{xx})$ with $\rho_{xx}$ the longitudinal resistivity and $e$ the electron charge). Consequently, the charge carriers in thick FeCl$_3$-FLG have a macroscopic mean free path as high as 0.6 $\mu m$ in 5L at room temperature. The outstanding electrical properties, e.g. lower $R_s$ than ITO and macroscopic mean free path, found in FeCl$_3$-FLGs thicker than 3L  are of fundamental interest for the development of novel electronic applications based on highly conductive materials.

Whether FeCl$_3$-FLGs can replace ITO in optoelectronic applications strongly depends on their optical properties. Surprisingly, our detailed study of the optical transmission in the visible wavelength range shows that while FeCl$_3$ intercalation improves significantly the electrical properties of graphene, it leaves the optical transparency nearly unchanged. Figures 3a and b show a comparison between the transmittance spectra of pristine FLG and FeCl$_3$-FLG. The transmittance values of pristine FLG at the wavelength of 550nm are in agreement with the expected values \cite{15transparency}, highlighted in Figure 3a, and with the results reported by other groups \cite{15transparency,17Roll}. Upon intercalation, the transmittance slightly decreases at low wavelengths, but it is still above 80 $\%$. In order to measure an accurate value of transmittance we fit it with a linear dependence on the number of layers for a statistical ensemble of flakes (Figure 3d). This results in similar extinction coefficients per layer for pristine FLG ($\approx 2.4$\textpm$0.1 \%$) and for FeCl$_3$-FLG ($\approx 2.6$\textpm$0.1\%$), see Figure 3d. For wavelengths longer than 550nm we observe an increase in the optical transparency of FeCl$_3$-FLG. This is a significant advantage of our material compared to ITO whose transparency decreases for wavelengths longer than 600nm \cite{REV2}. This property will provide useful applications that require conductive electrodes which are transparent both in visible and near infrared range. For instance, FeCl$_3$-FLG transparent electrodes could be used for solar cells to harvest energy over an extended  wavelength range as compared to ITO-based devices, or for electromagnetic shielding in infrared.

\begin{figure}[!h] 
\centering
\includegraphics[width=1\textwidth]{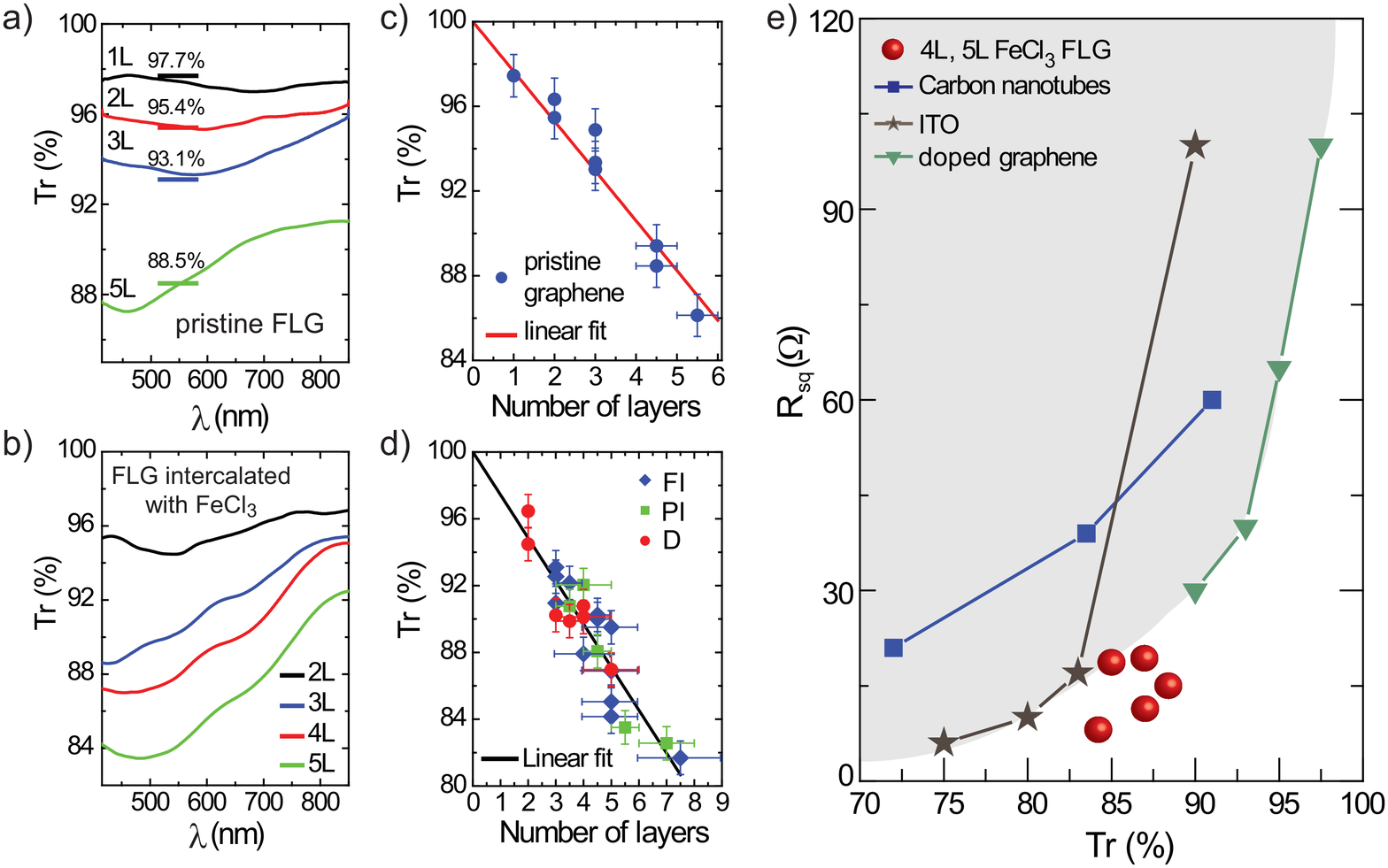}
\noindent{\caption{Panels a) and b) show the transmittance spectra of pristine FLG and FeCl$_3$-FLG, respectively. The horizontal lines in b) are the corresponding transmittances at the wavelength of $550nm$ reported in the literature \cite{15transparency,17Roll}. c) Transmittance at $550nm$ for pristine FLG as a function of the number of layers. The red line is a linear fit, which gives the extinction coefficient of $2.35$\textpm$0.1\%$ per layer. d) Transmittance at $550nm$ for fully intercalated FeCl$_3$-FLG (FI), partially intercalated FeCl$_3$-FLG (PI) and doped FeCl$_3$-FLG (D) as a function of the number of layers. The black line is a linear fit with the extinction coefficient of $(2.6$\textpm$0.1)\%$ per layer. e) Square resistance \textit{versus} transmittance at $550nm$ for 4L and 5L FeCl$_3$-FLG (from these experiments), ITO (from \cite{7metall2}), carbon-nanotube films (from ref \cite{4tube1}) and doped graphene materials (from ref \cite{17Roll}). FeCl$_3$-FLG outperform the current limit of transparent conductors, which is indicated by the grey area.} \label{fig3}}
\end{figure}

The high transparency observed in FeCl$_3$-FLG complemented by their remarkable electrical properties make these materials valuable candidates for transparent conductors. However, to replace ITO in optoelectronic applications, it is generally agreed that materials must (at least) have the properties of commercially available ITO (R$_{s}$=10 $\Omega/\Box $ and Tr=85$\%$ \cite{27limits}). Figure 3e compares R$_{s}$ \textit{vs.} Tr of FeCl$_3$-FLG materials with ITO \cite{7metall2}, and other promising carbon-based candidates to replace ITO such as carbon-nanotube films \cite{4tube1} and doped graphene materials \cite{17Roll}. It is apparent that R$_{s}$ and Tr of FeCl$_3$-FLGs outperform the current limits of ITO and of the best values reported so far for doped graphene \cite{17Roll}. Therefore, the outstandingly high electrical conductivity and optical transparency make FeCl$_3$-FLG materials the best transparent conductors for optoelectronic devices.

Finally, an important characteristic required by a transparent conductor is its stability upon exposure to air. In principle FLGs could be intercalated with a large variety of molecules, similar to the graphite intercalation compounds (GIC) \cite{20Dresselhaus}. However, most of the GIC are unstable in air, with donor compounds being easily oxidized and acceptors being easily desorbed. Therefore we studied the stability in air of FeCl$_3$-FLG by performing Raman measurements as a function of time. We found that the Raman spectra of FeCl$_3$-FLG samples show no appreciable changes on a time scale of up to one year (see Supporting Information). This property has important implications for the utilization of these materials as transparent conductors in practical applications such as displays and photovoltaic devices.

In conclusion, we demonstrate novel transparent conductors based on few layer graphene intercalated with ferric chloride with an outstandingly high electrical conductivity and optical transparency. We show that upon intercalation a record low sheet resistance of 8.8 $\Omega/\Box$ is attained together with an optical transmittance higher than 84$\%$ in the visible range. These parameters outperform the best values of ITO and of other carbon-based materials. The FeCl$_3$-FLGs materials are relatively inexpensive to make and they are easily scalable to industrial production of large area electrodes. Contrary to the numerous chemical species that can be intercalated into graphite (more than hundred \cite{20Dresselhaus}), many of which are unstable in air, we found that FeCl$_3$-FLGs are air stable on a timescale of at least one year. Other air stable graphite intercalated compounds can only by synthesized in the presence of Chlorine gas \cite{20Dresselhaus}, which is highly toxic. On the contrary, here we demonstrate that the intercalation of FLG with FeCl$_3$ is easily achieved without the need of using Chlorine gas, which ensures an environmental friendly industrial processing. Furthermore, the low intercalation temperature (360\textcelsius{}) required in the processing allows the use of a wide range of transparent flexible substrates which are compatible with existing transparent electronic technologies. These technological advantages combined with the unique electro-optical properties found in FeCl$_3$-FLG make these materials a valuable alternative to ITO in optoelectronics.\\

\noindent\textbf{Experimental section}

\noindent\textbf{Sample fabrication.}
The intercalation process with FeCl$_3$ is performed in vacuum. Both anhydrous FeCl$_3$ powder and the substrate with exfoliated FLG are positioned in different zones inside a glass tube. The tube is pumped down to $2\times10^{-4}mbar$ at room temperature for 1 hour to reduce the contamination by water molecules. Subsequently, the FLG and the powder are heated for 7.5 hours at 360\textcelsius{} and 310\textcelsius{}, respectively. A heating rate of 10\textcelsius{}/min is used during the warming and cooling of the two zones. Ohmic contacts are fabricated on FeCl$_3$-FLG by means of electron-beam lithography and lift-off of thermally evaporated chrome/gold bilayer (5/50 nm). We have fabricated FeCl$_3$-FLG on both SiO$_2$/Si and glass substrates and we found no significant differences in their transport properties.

\noindent\textbf{Raman measurements.} Raman spectra are collected in ambient air and at room temperature with a Renishaw spectrometer. An excitation laser with a wavelength of 532 nm, focused to a spot size of 1.5 $\mu$m diameter and a $\times$100 objective lens are used. To avoid sample damage or laser induced heating, the incident power is kept at 5 mW.

\noindent\textbf{Electrical measurements.} The longitudinal and the Hall resistances are studied in a 4-probe configuration by applying an a.c. current bias and measuring the resulting longitudinal and transversal voltages with a lock-in amplifier. The excitation current is varied to ensure that the energy range where electrical transport takes place is smaller than the energy range associated to the temperature of the electrons. This prevents heating of the electrons and the occurrence of nonequilibrium effects. Since conductances are additive, the analysis of Shubnikov-de Haas oscillations is performed on the longitudinal conductivity $\sigma_{xx}=\rho_{xx}/[\rho_{xx}^{2}+\rho_{xy}^{2}]$ (with $\rho_{xx}$ and $\rho_{xy}$ the longitudinal and transversal resistivity, respectively). The low frequency magneto-conductivity oscillations shown in Figure 2d are obtained by averaging out the high frequency oscillations, whereas to obtain the high frequency oscillations shown in Figure 2e we subtract the low frequency oscillations from the longitudinal conductivity.

\noindent\textbf{Transmission measurements.}
The transmission of pristine FLG and FeCl$_3$-FLG is characterized by measuring the bright-field transmission spectra. A system based on an inverted optical microscope (Nikon Eclipse TE2000-U) combined with a spectrometer and CCD camera (Princeton Instruments, SpectraPro 2500i) is used to acquire data. White light from a tungsten filament lamp is used to illuminate the samples and, after passing through the sample, is collected by a dry Nikon lens (S Plan Fluor ELWD) $\times$40 of NA 0.60. A slit width of 50 $\mu$m is used for the spectrometer, yielding a spectral resolution $<$ 1 nm for the measurements. In the spectrometer the dispersed light is projected onto the 1024 by 256 lines of the CCD camera. Data from the camera are extracted to give the transmission spectra of the flake, or part of it, the data being normalized to the signal obtained through a region of bare substrate. For the visually uniform parts of the flakes, spectra are averaged along several lines of the CCD camera to improve the signal-to-noise ratio.\\

\noindent\textbf{Supporting information.}

\noindent Supporting information is available from the Wiley Online Library (http://onlinelibrary.wiley.com/doi/10.1002/adma.201200489/suppinfo) or from the author\\

\noindent\textbf{Acknowledgements.}

\noindent We acknowledge Paul Wilkins for technical support. This work was financially supported by EPSRC (Grants No. EP/G036101/1 and no. EP/J000396/1) and from the Royal Society (Research Grants 2010/R2 no. SH-05052 and 2011/R1 no. SH-05323).\\

\end{document}